\documentclass[prl,twocolumn,superscriptaddress,showpacs]{revtex4}

\usepackage[T1]{fontenc}
\usepackage{times}
\usepackage{color,graphicx}
\usepackage[dvipdfmx,colorlinks=true]{hyperref}
\usepackage{amsmath,amssymb,amsthm}

\newcommand\ket[1]{\ensuremath{|#1\rangle}}
\newcommand\bra[1]{\ensuremath{\langle#1|}}
\newcommand\iprod[2]{\ensuremath{\langle#1|#2\rangle}}
\newcommand\oprod[2]{\ensuremath{|#1\rangle\langle#2|}}

\newcommand\supp{\mathop{\rm supp}}

\newtheorem{theorem}{Theorem}

\begin{document}

%% End-Of-Header

\title{No-go Theorem for One-way Quantum Computing on Naturally
  Occurring Two-level Systems}

\author{Jianxin Chen}%
\affiliation{Department of Computer Science and Technology, Tsinghua
  National Laboratory for Information Science and Technology, Tsinghua
  University, Beijing, China}%
\author{Xie Chen}%
\affiliation{Department of Physics, Massachusetts Institute of
  Technology, Cambridge, Massachusetts, USA}%
\author{Runyao Duan}%
\affiliation{Centre for Quantum Computation and Intelligent Systems
  (QCIS), Faculty of Engineering and Information Technology,
  University of Technology, Sydney, New South Wales, Australia}%
\affiliation{Department of Computer Science and Technology, Tsinghua
  National Laboratory for Information Science and Technology, Tsinghua
  University, Beijing, China}%
\author{Zhengfeng Ji}%
\affiliation{Perimeter Institute for Theoretical Physics, Waterloo,
  Ontario, Canada}%
\affiliation{State Key Laboratory of Computer Science, Institute of
  Software, Chinese Academy of Sciences, Beijing, China}%
\author{Bei Zeng}%
\affiliation{Institute for Quantum Computing and Department of
  Combinatorics and Optimization, University of Waterloo, Waterloo,
  Ontario, Canada}%

\begin{abstract}
  One-way quantum computing achieves the full power of quantum
  computation by performing single particle measurements on some
  many-body entangled state, known as the resource state. As single
  particle measurements are relatively easy to implement, the
  preparation of the resource state becomes a crucial task. An
  appealing approach is simply to cool a strongly correlated quantum
  many-body system to its ground state. In addition to requiring the
  ground state of the system to be universal for one-way quantum
  computing, we also want the Hamiltonian to have non-degenerate
  ground state protected by a fixed energy gap, to involve only
  two-body interactions, and to be frustration-free so that
  measurements in the course of the computation leave the remaining
  particles in the ground space. Recently, significant efforts have
  been made to the search of resource states that appear naturally as
  ground states in spin lattice systems. The approach is proved to be
  successful in spin-$\frac{5}{2}$ and spin-$\frac{3}{2}$ systems.
  Yet, it remains an open question whether there could be such a
  natural resource state in a spin-$\frac{1}{2}$, i.e., qubit system.
  Here, we give a negative answer to this question by proving that it
  is impossible for a genuinely entangled qubit states to be a
  non-degenerate ground state of any two-body frustration-free
  Hamiltonian. What is more, we prove that every spin-$\frac{1}{2}$
  frustration-free Hamiltonian with two-body interaction always has a
  ground state that is a product of single- or two-qubit states, a
  stronger result that is interesting independent of the context of
  one-way quantum computing.
\end{abstract}

\date{Apr. 21, 2010}


\maketitle

%-----------------------------------------------------------------------------%
% Motivation:
%-----------------------------------------------------------------------------%

Quantum computers are distinct from classical ones, not only in that
they can solve hard problems that are intractable on classical
computers, factoring large numbers for example~\cite{Sho99}, but also
in that they can be implemented in architectures such as one-way
quantum computing~\cite{RB01,RB02,WRR+05} that have no evident
classical analogues at all. Unlike the quantum circuit
model~\cite{Deu89,Yao93,BBC+95} which employs entangling gates during
the computation, one-way quantum computation requires only single
particle measurements on some prepared entangled state, also known as
the resource state. This new quantum computation scheme sheds light on
the role of entanglement in quantum computation and provides possible
advantages in physical implementation of quantum computers. Moreover,
from the theoretical computer science perspective, although one-way
quantum computations are polynomial time equivalent to the unitary
circuit model, they may have advantages over the circuit model in
terms of parallelisability~\cite{RB02,Joz05,BK09}. For example, the
quantum Fourier transform~\cite{NC00}, the key quantum part of Shor's
factoring algorithm, is approximately implementable in constant depth
in the one-way model~\cite{BKP09}. All these nice facts about the
one-way computing model make it a worthy topic to pursue both
theoretically~\cite{RBB03,GE07,GESP07,GE08,DKP07,Pop07,BBD+09,CDJZ10}
and experimentally~\cite{WRR+05}.

Quantum entanglement is believed to be a necessary ingredient of
quantum computation~\cite{SDV06,Vid03}; yet entangling operations used
in the unitary circuit model that generate and process quantum
entanglement are hard to implement on large scale systems.
Entanglement is also essential in one-way quantum
computing~\cite{NMDB06,NDVB07,GFE09,BMW09}. However, the entanglement
used in a one-way quantum computer is cleanly separated in the initial
preparation step from the whole computation. Moreover, it usually has
a regular structure and is independent of the computation problem and
inputs. This allows us to focus on the preparation of some specific
entangled resource state.

An appealing idea is to obtain the resource state in some strongly
correlated quantum many-body system at low temperature. This approach
requires the resource state to be the non-degenerate ground state of
some gapped Hamiltonian, which involves only two-body nearest-neighbor
interactions. In this way, the resource state can be effectively
created via cooling, and the procedure is robust against thermal
noises. The Hamiltonian also needs to be frustration free, that is,
the ground state minimizes the energy of each local term of the
Hamiltonian simultaneously, so that measurements in the course of the
computation leave the remaining particles in the ground space.

The canonical resource state for one-way quantum computing, known as
the cluster state~\cite{RB01}, does not naturally occur as a ground
state of a physical system~\cite{Nie06}. As a result, there has been
significant efforts to identify alternative resource states that
appear naturally as ground states in spin
lattices~\cite{GE07,GESP07,BM08,CZG+09,CMDB10}. In Ref.~\cite{CZG+09},
a natural resource state called triCluster is found in a
spin-$\frac{5}{2}$ system. And very recently, a two-body
spin-$\frac{3}{2}$ Hamiltonian from a quantum magnet is found, whose
unique ground state is also a universal resource state for one-way
quantum computing~\cite{CMDB10}. As two-level systems are more widely
available in practice than higher-level systems, it is natural to ask
whether there exists a universal resource state in spin-$\frac{1}{2}$
(qubit) system that naturally occurs.

In this letter, however, we show that it is not the case. Namely, a
genuinely entangled qubit states cannot be a non-degenerate ground
state of any two-body frustration-free Hamiltonian $H$, as there
is always a product of single qubit states in the ground space of $H$.
This indicates that one-way computing with naturally occurring
resource states cannot be done with qubits. Therefore, the best one
can hope for is to find natural resource state in spin-$1$ systems,
the existence of which remains an open question.

With a similar argument, we show that any two-body frustration-free
Hamiltonian has a ground state that is a product of single- or
two-qubit states. This leads to deeper understandings of the
relationship between frustration in the Hamiltonian and entanglement
in the ground state for qubit systems.

It is worth noting that our discussion is also closely related to a
problem in quantum computational complexity theory, the quantum analog
of $2$-Satisfiability (abbreviated as Quantum $2$-SAT~\cite{Bra06}).
We will discussion the relation in detail in the next section.

%-----------------------------------------------------------------------------%
% FF Hamiltonian:
%-----------------------------------------------------------------------------%

{\em The frustration-free Hamiltonian.---\/} We start our proof by
assuming that there does exist such a naturally occurring state of $n$
qubits, denoted by $\ket{\Psi}$. We also assume for simplicity that
the state is genuinely entangled, meaning that it is not a product
state with respect to any bi-partition of the $n$-qubit system.

Given any density matrix $\rho$, we define its support $\supp(\rho)$
to be the subspace spanned by the eigenvectors of non-zero eigenvalues
of $\rho$. For any two qubits $i,j$, the two-particle reduced density
matrix of these two qubits of state $\ket{\Psi}$ is denoted as
$\rho_{ij}$.

The state $\ket{\Psi}$ gives rise to a two-body frustration-free
Hamiltonian $H_\Psi$ that has $\ket{\Psi}$ in its ground space, and at
the same time, has the smallest possible ground space in a sense
formalized below. In fact, the Hamiltonian can be chosen, without loss
of generality, to be the sum of projections $\Pi_{ij}$ onto the
orthogonal space of the $\supp(\rho_{ij})$, that is,
\begin{equation}
\label{eq:FFH}
H_\Psi = \sum_{ij}\,\Pi_{ij}. 
\end{equation}
As $H_\Psi$ is constructed from state $\ket{\Psi}$, we call it the
two-body frustration-free Hamiltonian of $\ket{\Psi}$

Clearly $H_\Psi$ is two-body and frustration-free, and $\ket{\Psi}$ is
a ground state of $H_\Psi$ with energy $0$. Note that the ground space
of $H_\Psi$ is given by
\begin{equation}
\mathcal{S}(\ket{\Psi}) = \bigcap_{ij} \supp(\rho_{ij}\otimes I_{\bar{ij}}),
\end{equation}
where $I_{\bar{ij}}$ is the identity operator on qubits other than
$i,j$.

Generally, a frustration-free Hamiltonian $H$ needs not to be a summation
of projections. However, we can always find one whose local terms are
indeed projections and has the same ground space as $H$. Therefore, we
only consider frustration-free Hamiltonian that are summation of
projections in this paper. It is not hard to see that any two-body
frustration-free Hamiltonian $H'$ that has $\ket{\Psi}$ as a ground
state also contains $\mathcal{S}(\ket{\Psi})$ in its ground space. In
other words $H_\Psi$ has the smallest possible ground space among all
frustration-free Hamiltonians having $\ket{\Psi}$ as a ground state.

There is a natural correspondence between a two-body frustration-free
Hamiltonian $H$ and the Quantum $2$-SAT problem. Classically, a
$2$-SAT problem asks whether a logical expression in the conjunctive
normal form with two variables per clause, e.g. $(x_0\lor x_1) \land
(\lnot x_1 \lor x_2) \land (x_2 \lor \lnot x_0)$, is satisfiable or
not, where $x_i$ are Boolean variables and $\land,\lor,\lnot$ are
logical AND, OR, NOT operations. There is a well-known polynomial time
classical algorithm that solves $2$-SAT while the related $3$-SAT
problem is believed to be much harder (NP-Complete~\cite{NP}). In
Ref.~\cite{Bra06}, it was proved that the quantum analog of the
$2$-SAT problem, which asks whether a set of projections on two-qubit
subsystems has a simultaneous ground state, is also efficiently
solvable on a classical computer. It was also shown there that Quantum
$4$-SAT is one of the hardest problems in $\text{QMA}_1$ (a quantum
analog of NP~\cite{NP}), meaning that it is probably hard even for
quantum computers. The relation between a frustration-free Hamiltonian
and its corresponding quantum SAT problem is evident. The Hamiltonian
$H$ is indeed frustration-free, thereby having $0$ ground energy, if
and only if the quantum SAT problem defined by the set of projections
in the Hamiltonian $H$ is satisfiable. In the case of two-body
Hamiltonian $H_\Psi$, the corresponding Quantum $2$-SAT problem is
defined by $\Pi_{ij}$'s. If for each term $\Pi_{ij}$, the rank of it
is either $0$ or $1$, the corresponding Quantum $2$-SAT problem is
called homogeneous~\cite{Bra06}, a concept that will be used in the
following.

Now we go back to the Hamiltonian problem and show that $\ket{\Psi}$
cannot be a unique ground state of any two-body frustration-free
Hamiltonian by proving the following theorem.

\begin{theorem}
  Given an $n$-qubit state $\ket{\Psi}$ that is genuinely entangled
  and any two-body frustration-free Hamiltonian $H$ having
  $\ket{\Psi}$ as its ground state. There always exists a product
  state of single qubits also in the ground space of $H$ for $n \ge
  3$.
\end{theorem}

As $H_\Psi$ has the smallest ground space, we only need to prove the
theorem for $H_\Psi$ instead of the general $H$. Also, it is
equivalent to prove that $\mathcal{S}(\ket{\Psi})$ is of dimension at
least $2$ and contains a product state of single qubits.

%-----------------------------------------------------------------------------%
% Proof:
%-----------------------------------------------------------------------------%

{\em Proof of the theorem.---\/} We prove this theorem by induction.
Before doing so, we examine the following fact. Let $\ket{\Psi}$ and
$\ket{\Phi}$ be two $n$-qubit states that can be transformed into each
other by invertible local operations. That is, there are $2\times 2$
non-singular linear operators $L_1,\cdots, L_n$, such that $\ket{\Psi}
= \mathcal{L} \ket{\Phi}$, where $\mathcal{L}=L_1\otimes \cdots\otimes
L_n$. This is equivalently to saying that $\ket{\Psi}$ and
$\ket{\Phi}$ can be transformed to each other via stochastic local
operation and classical communication (SLOCC)~\cite{BPR+00,DVC00}.
Noticing the fact~\cite{BMR09} that $\ket{\Psi}$ is a ground state of
$H$ if and only if $\ket{\Phi}$ is a ground state of
\begin{equation}
  H' = \sum_{ij} (L_i \otimes L_j)^\dagger \Pi_{ij} (L_i
\otimes L_j),
\end{equation}
and the trivial fact that $\mathcal{L}$ maps product states to product
states, we only need to discuss states that are representatives of
equivalent classes induced by such local transforms $\mathcal{L}$.
Equivalently, it suffices to consider SLOCC equivalent classes.

For three-qubit genuinely entangled states, there are only two
different SLOCC equivalent classes~\cite{DVC00}, represented by the
$\ket{W}$ and $\ket{GHZ}$ respectively where $\ket{W} = (\ket{001} +
\ket{010} + \ket{100}) / \sqrt{3}$, and $\ket{GHZ} = (\ket{000} +
\ket{111}) / \sqrt{2}$. For the $\ket{W}$ state, one has
\begin{equation}
\mathcal{S}(\ket{W}) = \mathop{\rm span} \{ \ket{W}, \ket{000} \},
\end{equation}
therefore the product state $\ket{000}$ is in the ground space. 
While for $\ket{GHZ}$, 
\begin{equation}
\mathcal{S}(\ket{GHZ}) = \mathop{\rm span} \{ \ket{000}, \ket{111} \},
\end{equation}
both product states $\ket{000}$ and $\ket{111}$ are in the ground
space. This then proves the theorem for the three-qubit case.

Now we proceed to the four-qubit case. Note that $\ket{\Psi}$ is
genuinely entangled, all $\rho_{ij}$ must be of rank at least $2$,
i.e. the dimension of the $\supp(\rho_{ij})$ is at least $2$ and the
rank of $\Pi_{ij}$ is at most $2$. We will discuss two cases here.

{\it Case 1.} If for some pair of qubits, say $(3,4)$, the rank of
their reduced density matrix $\rho_{34}$ is $2$, then the pair of
qubits $3,4$ can be encoded as a single qubit. Therefore, we can
reduce our problem to a similar one of smaller system size.

To be more precise, suppose $\rho_{34}$ is supported on two orthogonal
states $\ket{\psi_0}_{34}$ and $\ket{\psi_1}_{34}$. Define an isometry
\begin{equation}
  V: \ket{0}_{3'} \rightarrow \ket{\psi_0}_{34},
  \ket{1}_{3'} \rightarrow \ket{\psi_1}_{34},
\end{equation}
which maps a single qubit to two qubits. That is, we have used qubit
$3'$ to encode the two qubits $3,4$. Define $\ket{\Phi}=V^\dagger
\ket{\Psi}$, so $\ket{\Psi}$ is a ground state of $H$ if and only if
$\ket{\Phi}$ is a ground state of $H'=V^\dagger HV$. One can easily
verify that $H'$ is still a two-body frustration-free Hamiltonian and
$\ket{\Phi}$ is a genuinely entangled state of $3$ qubits. This
reduces to a case already proved and there is product state
$\ket{\alpha_1}\otimes\ket{\alpha_2}\otimes\ket{\alpha_{3'}}$ which is
also a ground state of $H'$.

Let $\ket{\beta_{34}}$ be $V \ket{\alpha_{3'}}$, a two-qubit state of
qubits $3,4$. If it is a product state, then we are done. If it is
entangled, as $\rho_{34}$ is supported on a $2$-dimensional space,
there always exists a product state $\ket{\beta_3} \otimes
\ket{\beta_4} \in \supp(\rho_{34})$~\cite{Par04}. Consider now the
bi-partition between qubits $1,2$ and qubits $3,4$. As
$\ket{\beta_{34}}$ is entangled, any projection term that concerns two
qubits from different partitions will have trivial constraints on
qubits $3$ and $4$. Therefore, the product state $\ket{\alpha_1}
\otimes \ket{\alpha_2} \otimes \ket{\beta_{3}} \otimes \ket{\beta_4}$
is also a ground state of $H$.

\begin{figure}[htbp]
  \centering
  \includegraphics{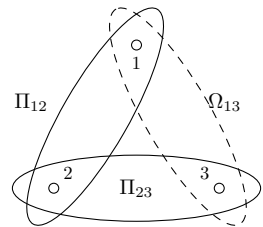}
  \caption{An illustration of completion procedure}
  \label{fig:completion}
\end{figure}

{\it Case 2.} If all of the $\supp(\rho_{ij})$ are of rank $3$ or $4$,
we employ the homogeneous $2$-SAT and completion techniques in
Ref.~\cite{Bra06} to finish the proof. The completion procedure adds
possibly new projection terms to the frustration-free Hamiltonian
without changing the ground space. For any three qubits, say $1,2,3$,
the procedure takes two rank-$1$ Hamiltonian terms say $\Pi_{12}$ and
$\Pi_{23}$, and generates a possibly new constraint $\Omega_{13}$. See
Fig.~\ref{fig:completion} for an illustration. We briefly review the
specific rule for obtaining $\Omega_{13}$ from $\Pi_{12}$ and
$\Pi_{23}$; and refer the interested readers to Ref.~\cite{Bra06} for
the proof and details. Let $\Pi_{12} = \oprod{\phi}{\phi}$, $\Pi_{23}
= \oprod{\theta}{\theta}$, and $\Omega_{13} = \oprod{\omega}{\omega}$,
where $\ket{\phi},\ket{\theta},\ket{\omega}$ are two-qubit pure states.
Denote, for example, $\phi_{\alpha,\beta}$ as the amplitude
$\iprod{\alpha,\beta}{\phi}$. Then relation is given by
$\omega_{\alpha,\gamma} = \phi_{\alpha,\beta} \epsilon_{\beta,\delta}
\theta_{\delta,\gamma},$ where $\epsilon = \oprod{0}{1}-\oprod{1}{0}$
and the summation of repeated indices is implicit~\cite{Bra06}.

The key point here is that the construction of $H_\Psi$ guarantees
that no new constraint could ever been added during the completion
procedure. Therefore, $H_\Psi$ corresponds to a Quantum $2$-SAT that
satisfies all the conditions (homogeneous and completed) in Lemma 2 of
Ref.~\cite{Bra06} and it follows that there is a product of
single-qubit states in the ground space of $H_\Psi$.

This proves the theorem for the four-qubit case and the general
$n$-qubit case can be proved by the same induction.

%-----------------------------------------------------------------------------%
% Entanglement and Frustration:
%-----------------------------------------------------------------------------%

{\em Entanglement versus frustration for qubit system.---\/} Our main
result is a no-go theorem for one-way quantum computing, which says
that in order to do one-way quantum computing with a natural ground
state, one has to go to higher dimensional particle systems other than
two-level systems. Interestingly, a similar argument also gives a
better understanding for the relationship of entanglement and
frustration for qubit systems. We can modify our method to show the
following result.

\begin{theorem}
  For any two-body frustration-free Hamiltonian $H$ of a qubit system,
  there always exists a ground state, which is a product of single- or
  two-qubit states.
\end{theorem}

To see this, first note that if any local term $\Pi_{ab}$ in $H$ is of
rank $3$, then the corresponding ground state can only be of the form
$(I-\Pi_{ab})\otimes\ket{\Psi'}\bra{\Psi'}$, where $(I-\Pi_{ab})$ is
of rank $1$ and $\ket{\Psi'}$ is a state of remaining qubits.
Secondly, as the Hamiltonian is now of a more arbitrary form, we may
not automatically have the completion property as in $H_\Psi$. Yet, it
is easy to overcome this problem by the use of completion procedure as
in Ref.~\cite{Bra06} and we omit the details for simplicity.

This theorem indicates that frustration is a necessary condition for
genuine many-body ground state entanglement in a natural qubit system
with non-degenerate ground state.

In the language of Quantum $2$-SAT, the above theorem states that if a
Quantum $2$-SAT is satisfiable, there will be a ground state that is
product of single- or two-qubit states. This is a much simpler form
than the recursive construction in Ref.~\cite{Bra06}.

If we further require some symmetry of the Hamiltonian, say, certain
kind of translational invariance, there could be only two phases for a
non-degenerate frustration-free system with qubits at zero
temperature: one is a product state phase, and the other is a dimer
phase~\cite{Sac08}. This relationship of entanglement and frustration
is not true in a spin-$1$ (qutrit) system. For instance, the famous
Affleck-Kennedy-Lieb-Tasaki (AKLT) state~\cite{AKLT87} is a
non-degenerate ground state of a two-body frustration-free Hamiltonian
on a chain. Interestingly, the AKLT state and some of its variants on
a chain are indeed powerful enough to process single qubit information
in the one-way quantum computing model~\cite{GE07,BM08,CDJZ10,LKZ+10}.

%-----------------------------------------------------------------------------%
% Summary:
%-----------------------------------------------------------------------------%

{\em Summary and Discussion.---\/} We have shown that it is impossible
for a genuinely entangled qubit state to be a unique ground state of
any two-body frustration-free Hamiltonian $H$, because there is always
a product state of single qubits also in the ground space of $H$. This
indicates that one-way computing cannot be done on naturally occurring
qubit systems. Furthermore, we use similar technique to prove that
every spin-$\frac{1}{2}$ frustration-free Hamiltonian with two-body
interaction always has a ground state that is a product of single- or
two-qubit states. These results are strong in the sense that they are
independent of the lattice structure, and therefore valid for any
lattice geometry with natural nearest-neighbour interactions in the
Hamiltonian.

A direct consequence also follows for condensed matter theory. Namely,
without degeneracy, there is no genuine many-body entanglement in a
ground state of a spin-$\frac{1}{2}$ frustration-free Hamiltonian with
two-body interaction. This is not the case for frustration-free higher
spin systems or spin-$\frac{1}{2}$ systems with more than two-body
interactions. These observations are also closely related to the study
of quantum computational complexity theory, which shows that Quantum
$2$-SAT is easy, but Quantum $2$-SAT with large enough local
dimensions or Quantum $3$-SAT might be much more
difficult~\cite{Bra06,ER08}. Our result also simplifies the structure
of the solution space of Quantum $2$-SAT given in Ref.~\cite{Bra06}.
However, a full characterization of the solution-space structure needs
further investigation. We hope that our result helps in further
investigations of local Hamiltonian problems and in linking the fields
of condensed matter, quantum information and computer science.

\begin{acknowledgments}
  We thank S. Bravyi and X.-G. Wen for valuable discussions. RD is
  partly supported by QCIS, University of Technology, Sydney, and the
  NSF of China (Grant Nos. 60736011 and 60702080). ZJ acknowledges
  support from NSF of China (Grant Nos. 60736011 and 60721061); his
  research at Perimeter Institute is supported by the Government of
  Canada through Industry Canada and by the Province of Ontario
  through the Ministry of Research \& Innovation. BZ is supported by
  NSERC and QuantumWorks.
\end{acknowledgments}

\bibliography{qubits}

\begin{thebibliography}{10}
\bibitem{NP}%
  \BibitemOpen
  \bibinfo {note} {The class NP is the collection of problems that can be
  verified in polynomial time on a classical computer. A problem is NP-Complete
  if any problems in NP can be reduced to it in polynomial time.}%
  \bibAnnoteFile{Stop}{NP}%
\end{thebibliography}

%% Start-Of-Trailer

\end{document}